\documentclass[twocolumn]{aastex63}

\usepackage{xcolor}
\usepackage{graphicx, natbib, color, bm, amsmath, epsfig, mathrsfs, xfrac,
xspace}
\usepackage{ulem}
\usepackage{acronym}

\definecolor{orange}{rgb}{1.        ,  0.54,  0}
\definecolor{remove}{rgb}{0.5        ,  0.5,  0.5}

\newcommand{\grad}{\vec{\nabla}}
\newcommand{\divg}{\vec{\nabla} \cdot}

\newcommand{\D}{\mathrm{d}}

\newcommand{\fsh}{f_{\text{sh.}}}
\newcommand{\Max}{\mathcal{M}_t}

\newcommand{\mach}{\mathcal{M}_{\rm 1D}}
\newcommand{\machM}{\mathcal{M}_{M}}

\newcommand{\KE}{K}
\newcommand{\TE}{F}
\newcommand{\Enet}{E_{\rm net}}

\newcommand{\TEM}{\mathcal{F}}

\newcommand{\ke}{\kappa}
\newcommand{\te}{\varphi}

\begin{document}

\title{Statistics of Energy in Isothermal Supersonic Turbulence}

\begin{abstract}
Turbulence is a key process in many astrophysical systems.
In this work we explore the statistics of thermal and kinetic energy of isothermal, supersonic, turbulent gas. We develop analytic formulas for the PDF of thermal and kinetic energies and their joint PDF. We compare these analytical models with a suite of simulations with a fixed resolution of $1024^3$ cells across 4 different Mach numbers \mbox{(1, 2, 4, 8)} and three different driving patterns (compressive, mixed, solenoidal). We discover an interesting jump discontinuity in the thermal energy PDF, which carries onto the joint PDF.
\end{abstract}


\author[0000-0001-5372-6882] {Branislav Rabatin}
\affil{Department of Physics, Florida State University, Tallahassee, Florida}
\author[0000-0001-6661-2243]{David C. Collins}
\affil{Department of Physics, Florida State University, Tallahassee, Florida}

\section{Introduction}
\label{sec:intro}

Supersonic turbulence is a dynamical mechanism governing many aspects of galaxy evolution. Statistical modelling of turbulent gas offers insight into the dynamical evolution of the interstellar medium (ISM) and star formation. Supersonic motion is characteristic for star-forming regions, wherein the occurence of local collapse is triggered by large density fluctuations typical of supersonic flows (\citealt{Scalo04}, \citealt{MacLow04}) Turbulent molecular clouds in astrophysical simulations are commonly treated as isothermal. This expedient simplification is warranted due to the relatively constant temperature of the molecular clouds maintainted by their effective cooling rates.

In this study we set out to explore the statistical properties of thermal and kinetic energy involved in the dynamics of supersonic isothermal turbulence by describing the probability density function (PDF) of each component, as well as the joint statistics of the two. Through this analysis, further understanding of the statistical nature of isothermal supersonic turbulence is obtained. This focus opens further possibilities to model the complex dynamics of the ISM, extended to magnetized, self-gravitating turbulence with more realistic thermal evolution.

This is the third paper in a series.  The first paper
\citep{fsh23} explored the PDF of density.  This is treated as a lognormal in many works \citep[e.g.][]{Vazquez-Semadeni94}.  A lognormal is reasonable, as the density in supersonic isothermal turbulence can be treated as the result of a series of shocks, and an infinite number of random multiplications results in a lognormal distribution.  If one instead considers that a parcel of gas experiences only a small finite number of shocks, one obtains the Finite Shock Model (FSM). In \citet{fsh23} we show that FSM fits simulated data quite well, and naturally explains the variation with forcing parameter.  In the second paper \citep{rabatin23_correlations}, we examine the joint distribution of density and velocity, and develop an analytic model for their correlation.  Often assumed to be uncorrelated, we find a correlation that depends on Mach number, and can be described as a combination of differently weighted marginalized distributions.

This paper is organized into several parts. In Sec. \ref{sec:methods} we describe the simulation code, parameters, and numerical data analysis. In Sec. \ref{sec:overview} we review the dynamical quantities describing isothermally turbulent medium - density $\rho$, and velocity $\vec{v}$, as well as their statistical properties. In Sec. \ref{sec:thermal} we show, that the thermal energy density per unit volume of isothermal medium only depends on density. Therefore, its PDF, discussed in detail in Sec. \ref{sec:thermal_pdf}, can be built on our previously developed Finite Shock Model of density. In Sec. \ref{sec:kinetic} we focus on the kinetic energy density per unit volume. Properties of the joint PDF of thermal and kinetic energy are discussed in Sec. \ref{sec:joint}. We conclude in Sec. \ref{sec:conclusions}.
\section{Methods}
\label{sec:methods}

In this section we describe simulations that were used to verify our new model and present the methods used for statistical analysis.

\subsection{Simulations}

To test our theoretical results a set of turbulent, supersonic, isothermal simulations were performed using the hydrodynamic code {\sc Enzo} \citep{Bryan14}. The piecewise parabolic method \citep{Woodward84} is used to capture the shocks naturally occurring in a supersonically turbulent medium. The simulations are performed within a cube of unit length $L$, periodic in all directions, at a resolution of $1024^3$ cells, each having the same volume $\delta V = 1 / 1024^3$.

To maintain a steady, turbulent state, the Stochastic forcing module \citep{Federrath08} is used to drive the medium by adding energy at the large scale, at a rate proportional to the dissipation, $\dot{E} \sim M^3 / L$, where $M$ is the Mach number \citep{MacLow99,MacLow04}. The large-scale driving pattern is modelled after the \citet{Ornstein} process. In addition to the 1D r.m.s. Mach number $\mach$, each simulation is also characterized by the forcing mode $\xi$. The forcing mode determines the proportion of the solenoidal components within the forcing field; a purely solenoidal driving corresponds to $\xi = 1$, while $\xi = 0$ represents compressive driving.

For the purpose of extracting statistical properties from the simulations we define the turnover time scale $\tau$. The turnover scale is roughly equal to the turbulent crossing time $\tau_{\text{turb.}} = (L/2) / \mach$, where $L/2$ is the size of the driving pattern and $\mach$ is the 1D r.m.s. Mach number. Two frames, when taken at times separated by at least $\tau$ are said to be statistically uncorrelated. All simulations are run for 9 turnover scales, while the individual snapshots are taken with the step of $\tau / 10$. To capture the steady turbulent state, frames with $t < 2 \tau$ are discarded, leaving 71 snapshots available for statistical analysis. The 71 frames are analyzed together to provide a robust statistical ensemble. A similar approach is common in other works (\citet{Porter99}, \citet{Porter00}, \citet{Federrath10}, \citet{Federrath13}, \citet{Federrath21}).

Numerical simulations were performed using three forcing modes, compressive, mixed and solenoidal ($\xi = 0, 1/2, 1$) and four target Mach numbers, $\mach = 1, 2, 4, 8$. Simulation parameters and other extracted statistical quantities are summarized in Table \ref{tab:simpars}. The first column of the table identifies each simulation in the form of $\xi - M$, with forcing $\xi$ and target Mach number $M$. The actual 1D r.m.s. Mach number $\mach$ is listed in the second column, followed by the ratio of volume-weighted Mach number to mass-weighted Mach number squared, $\mathfrak X$. The third and last columns show the average thermal and kinetic energy, $\langle \TE \rangle = \mu_M$, $\langle \KE \rangle = \frac{3}{2} \machM^2$.

Speed of sound $c_s = \sqrt{p / \rho}$ and mean density $\rho_0 = M / V$ are simulation constants which will be throughout this work, for the sake of brevity, set to 1.

    \begin{table}
    \begin{center}
\begin{tabular}{| c || c | c || c | c |}
\hline
Sim. ($\xi-M$) & $\mach$ & $\mathfrak X$ & $\left\langle \TE \right\rangle$ & $\left\langle \KE \right\rangle$ \\ \hline \hline
$0-1$ & $0.977$ & $1.35$ & $0.910$ & $1.06$ \\
$0-2$ & $1.99$ & $1.38$ & $1.75$ & $4.31$ \\
$0-4$ & $3.92$ & $1.43$ & $2.49$ & $16.1$ \\
$0-8$ & $7.77$ & $1.41$ & $3.14$ & $64.1$ \\
$1/2-1$ & $0.999$ & $1.14$ & $0.191$ & $1.31$ \\
$1/2-2$ & $2.00$ & $1.21$ & $0.588$ & $4.99$ \\
$1/2-4$ & $3.98$ & $1.16$ & $1.03$ & $20.5$ \\
$1/2-8$ & $7.89$ & $1.14$ & $1.37$ & $81.8$ \\
$1-1$ & $0.993$ & $1.15$ & $0.160$ & $1.29$ \\
$1-2$ & $1.98$ & $1.16$ & $0.489$ & $5.08$ \\
$1-4$ & $3.71$ & $1.14$ & $0.840$ & $18.2$ \\
$1-8$ & $8.04$ & $1.13$ & $1.12$ & $86.1$ \\
\hline
\end{tabular}
\caption {Simulation parameters.  The first column denotes each simulation in the form of $\xi-M$ where $\xi$ is the forcing mode and $M$ is the nominal 1D r.m.s. Mach number. The second column lists the measured 1D r.m.s. Mach number. The third column represents $\mathfrak X = \langle v^2 \rangle / \langle \rho v^2 \rangle$, the ratio between the volume- and mass-weighted Mach numbers, squared. Last two columns show the average thermal and kinetic energy, $\langle \TE \rangle, \langle \KE \rangle$. The average thermal energy is related to the mass-weighted mean of log density, $\mu_M$ (see Sec. \ref{sec:thermal_general}), whereas the average kinetic energy is related to the mass-weighted Mach number, $\langle \KE \rangle = \frac{3}{2} \machM^2$.}
\label{tab:simpars}
\end{center}
\end{table}

\subsection{Statistical analysis}

Any random quantity $Q (\vec{x})$ defined within a simulation domain $V$ gives rise to its corresponding probability density function (PDF), which can be calculated as
\begin{equation}
    f_Q (q) = \frac{1}{V} \int_V \D^3 x \, \delta \big(q - Q (\vec{x}) \big)
    \label{eq:PDF_general}
\end{equation}
where $q$ is the corresponding random variable and $f_Q (q) \, \D q$ represents the probability of $Q$ occurring on the interval between $q$ and $q + \D q$.

Two random quantities $Q (\vec{x}), U (\vec{x})$, in addition to their individual statistics, also define the joint PDF, which can be found as follows
\begin{equation}
    f_{(Q, U)} (q, u) = \frac{1}{V} \int_V \D^3 x \, \delta \big(q - Q (\vec{x}) \big) \, \delta \big(u - U (\vec{x}) \big)
\end{equation}
with $q, u$ being the corresponding random variables. This approach is generalizable to an arbitrary number of random quantities.

Given a known PDF of $q$, $f_Q (q)$, and known relation between the two correpsonding random quantities $R = \phi (Q)$, the PDF of $R$, $f_R (r)$, can be obtained by the following simple random variable transformation
\begin{equation}
    f_R (r) = \int \D q \, f_Q (q) \, \delta \big(r - \phi (q) \big)
    \label{eq:1_1_transform}
\end{equation}
which naturally preserves the probability fraction, ${f_R (r) \, \D r = f_Q (q) \, \D q}$.

A known multivariate PDF of $q, u, \dots$ can be used to determine PDF of any transformed random variables depending on $q, u, \dots$ Specifically, PDFs for the cases of $2 \to 1$ variables, with the relation between the quantities given by $R = \phi (U, Q)$, and $2 \to 2$ variables, given by $R = \phi_1 (Q, U)$, $V = \phi_2 (Q, U)$, can be determined using the following formulas
\begin{align}
    \label{eq:2_1_transform}
    f_R (r) &= \int \D q \int \D u \, f_{(Q, U)} (q, u) \, \delta \big( r - \phi (q, u) \big) \\
    f_{(R, V)} (r, v) &= \int \D q \int \D u \, f_{(Q, U)} (q, u) \, \times \notag \\
    \label{eq:2_2_transform}
    &\hspace{65pt}\delta \big( r - \phi_1 (q, u) \big) \, \delta \big( v - \phi_2 (q, u) \big)
\end{align}

In this work, log density $s = \log \rho$ and speed $v$ will take the role of the baseline random variables with known statistical properties. Their dependent counterparts will be thermal energy $\TE (s) = s e^s$ and kinetic energy $\KE (s, v) = e^s v^2 / 2$. Equation \eqref{eq:1_1_transform} proves useful in determining the PDF of thermal energy in Sec. \ref{sec:thermal},  transformation \eqref{eq:2_1_transform} is used to obtain the PDF of kinetic energy in Sec. \ref{sec:kinetic} and the joint PDF of thermal and kinetic energy is obtained in Sec. \ref{sec:joint} using \eqref{eq:2_2_transform}.
\section{Isothermal turbulence: overview}
\label{sec:overview}


The dynamics of isothermal compressible gas in the inertial subrange, where both driving and dissipation can be neglected, is governed by the equation of continuity and Euler's equation
\begin{align}
    \label{eq:hydro_cont}
    \frac{\partial \rho}{\partial t} + \divg \left( \rho \vec{v} \, \right) &= 0 \\
    \label{eq:hydro_Euler}
    \rho \left( \frac{\partial \vec{v}}{\partial t} + \vec{v} \cdot \grad \vec{v} \right) &= - \grad p
\end{align}
where $\rho$ is density and $\vec{v}$ is velocity. The system of governing equations is closed by the isothermal equation of state, in which pressure is directly proportional to density
\begin{equation}
    p = c_s^2 \rho
\end{equation}
where $c_s$ is the speed of sound, conventionally set to 1 in all subsequent expressions.  

We will find it convenient to work with the log of density, $s = \log \rho/\rho_0$, where $\rho_0$ is the mean density.

Assuming given initial conditions at $t = 0$ and boundary conditions at all later times the equations determine the fields evolving over time and space. Periodic boundary conditions in a box with a side of length $L$ are used. Driving terms at largest scales and dissipation on the smallest scale given by the grid resolution balance out.

\subsection{Single-variable statistics}\label{sec:single_var}

In this section we summarize the statistical properties of the basic dynamical variables, density $\rho$ and speed $v$, as well as develop mathematical tools that will be useful later.

\subsubsection{Density Distribution: Finite Shock Model}

We consider a cloud of supersonic turbulence as an ensemble of shocks.
Each shock adjusts the pre-shock density by a multiplicative factor $m^2$ where $m = v / c_s$ is the local Mach number, drawn from the the velocity distribution, $f_v(v)$. The perfect log-normal behavior emerges as a consequence of the central limit theorem if a parcel of gas, on average, experiences a large amount of shocks $n$, each adjusting its density by a random multiplicative factor. In \citet{fsh23}, we improve this model by assuming that $n$ is finite and perhaps small.

We model $s$ as the sum of $n$ independent, identical events ${Y \sim \log m^2}$, normalized to zero mean and unit variance, as
\begin{equation}
    s = \frac{1}{\sqrt{n}} \sum_{i = 1}^n Y_i,
\end{equation}
which gives a charachteristic function of 
\begin{align}
\phi (\omega; n) \equiv \left\langle e^{i \omega s} \right\rangle = \left\langle e^{i \omega Y / \sqrt{n}} \right\rangle^n.
\end{align}
The distribution is then the inverse Fourier transform of the characteristic function,
\begin{equation}
	\fsh (s; n) = \int \limits_{- \infty}^{\infty} \frac{\D \omega}{2 \pi} e^{- i \omega s} \phi (\omega; n) .
	\label{eq_Fourier}
\end{equation}

The behavior can be captured to a great degree of accuracy by measuring two parameters from the ensemble; mean $\mu \equiv \langle s \rangle$ and variance $\sigma^2 \equiv \langle s^2 \rangle - \mu^2$. Values $\mu$, $\sigma$, $n$ and $\langle e^s \rangle$ are related as follows
\begin{equation}
    \langle e^s \rangle = e^\mu \phi (- i \sigma; n)
    \label{eq:fsh_exp_mean}
\end{equation}

Since $\langle e^s \rangle = \langle \rho / \rho_0 \rangle = 1$, mean $\mu$ and standard deviation $\sigma$ of log density uniquely determine the value for $n$ by way of \eqref{eq:fsh_exp_mean}. All other expected values involving density can also be calculated using the characteristic function, further details can be found in \citep{fsh23}.

\subsubsection{Velocity Distribution: Stretched Maxwell}
Due to the symmetries of the turbulent motion, the statistics of speed can be approximated by the Maxwellian distribution
\begin{equation}
    \mathcal M (v; \mach) = \frac{4 \pi v^2}{(2 \pi \mach^2)^{3/2}} \exp \left( - \frac{v^2}{2 \mach^2} \right),
\end{equation}
where $\mach = \sqrt{\langle v^2 \rangle / 3}$ is the 1D r.m.s. Mach number. As described in \citep{rabatin23_correlations},  the statistics of speed in our simulations can be described much more accurately by introducing a correction term of a higher power in $v$ within the exponential,
\begin{equation}
    \Max (v; \mach, b) \propto v^2 \exp \left[ - \frac{v^2}{2 a^2} \left( 1 - b + \frac{b v^2}{a^2} \right) \right]
    \label{eq:quartic_fit_VM}
\end{equation}
where the strength of the correction is adjusted by the sloping parameter $b \in [0,1]$ and $a$ is tuned so that the r.m.s. Mach number matches the desired value, $3 \mach^2 = \langle v^2 \rangle$. 
Further details can be found in \citep{rabatin23_correlations}.


\subsubsection{Velocity Distribution: Finite Shock Model}

We can also approximate the velocity distribution with FSM.
The density fluctuations are generated by random multiplicative factors drawn from the velocity distribution.  Similarly, the statistics of $v$ can be described in terms of FSM itself, with exactly one shock. Specifically, the PDF of the random variable $w \equiv \log v^2$ is given by
\begin{align}
    f_w (w; \mach, b) &= \Max^{(\text{log})} (w; \mach, b) \nonumber\\
    &\equiv e^{w / 2} \Max (e^{w / 2}; \mach, b)\nonumber \\
    &\approx \fsh (w; \mu_w, \sigma_w, n_w)
    \label{eq:max_fsh_approx}
\end{align}
where $\mu_w, \sigma_w, n_w=1$ are the mean, standard deviation and the effective number of shocks of $w$.
The mean and variance can be calculated analytically for $b = 0$
\begin{align}
    \mu_w (\mach, b = 0) &= 2 \log \mach + \Upsilon_0 \\
    \sigma_w (b = 0) &= \Sigma_0
\end{align}
with $\Upsilon_0 = 2 - \gamma - \log 2 \approx 0.730$ and $\Sigma_0 = \sqrt{\pi^2 / 2 - 4} \approx 0.967$ are numerical parameters related to the ideal Maxwellian distribution (here $\gamma$ refers to the Euler constant). Note, that while the mean of $\log v^2$ depends on the Mach number, the variance is scale free. 

The stretched Maxwellian can also be approximated by the Finite Shock Model, with modified parameters reflecting the change in mean and variance of $\log v^2$. The most extreme case when $b = 1$ can be again obtained analytically,
\begin{align}
    \mu_w (\mach, b = 1) &= 2 \log \mach + \Upsilon_1 \\
    \sigma_w (b = 1) &= \Sigma_1 \\
    n_w &\approx 1
\end{align}
where $\Upsilon_1 = \pi / 4 - \gamma / 2 + (1/2) \log 2 + \log \big[ \Gamma (7/4) / \Gamma (5 / 4) \big] \approx 0.857$, $\Sigma_1 = \sqrt{\pi^2 / 4 - 2 C} \approx 0.797$, with $C$ being the Catalan constant.

The approximation \eqref{eq:max_fsh_approx} can be used across the full range of parameters $b$ for the following interpolation between the cases for $b = 0$ and $b = 1$
\begin{align}
\label{eq:muw_approx}
    \mu_w (\mach, b) &\approx 2 \log \mach + 0.730 + \frac{2.170 b}{1 + 16.01 b^{0.82}} \\
\label{eq:sigmaw_approx}
    \sigma_w (b) &\approx 0.967 - \frac{2.349 b}{1 + 12.84 b^{0.78}} \\
    n_w &= 1
\end{align}

\subsubsection{Statistical weighting}\label{sec:weighting}

We will find it useful to compute distribution with different weighting.
Any PDF calculated from an ensemble via eq. \ref{eq:PDF_general} can be weighted by a non-negative quantity $W$,
\begin{equation}
    f^{(W)}_Q(q) = \frac{1}{W_{\text{net}}} \int_V d^3 x \, W(\vec{x}) \, \delta( q -
Q(\vec{x})),
\end{equation}
where $W_{\text{net}}$ is the total of $W$ within the ensemble. Alternative weighting of the same quantity can be useful in certain circumstances. In the context of this work, weighting by volume ($V$), mass ($M$) and kinetic energy ($\KE$) is utilized.

In addition to the log-density PDF, by default weighted by volume, the PDF weighted by mass and kinetic energy is required. In \citep{rabatin23_correlations} we find, that the mass- and kinetic energy-weighted instances can be related to the volume-weighted PDF, ${f^{(V)}_s (s) = \fsh (s; \mu, \sigma, n)}$, in the following way
\begin{align}
    f^{(M)}_s (s) &= e^s f^{(V)}_s (s) = e^s \fsh (s; \mu, \sigma, n) \\
    f^{(\KE)}_s (s) &= f^{(M)}_s (s + \log \mathfrak X) = \mathfrak X \, e^s \fsh (s; \mu - \log \mathfrak X, \sigma, n)
\end{align}
where $\mathfrak X = (\mach / \machM)^2$ is the square of the ratio between the 1D r.m.s. Mach number and its mass-weighted counterpart.

The mass-weighted PDF of speed becomes relevant to build the joint PDF of density and speed. It cannot be related to its volume-weighted counterpart, but can still be simply described by the stretched Maxwellian using its own set of parameters
\begin{equation}
    f^{(M)}_v (v) = \Max (v; \machM, b_M)
\end{equation}

\subsubsection{Convolutions of Finite Shock Models}
Two FSM distributions can be convolved together. This can be interpreted as a random variable $s = s_1 + s_2$ being the sum of two independent random variables. The result of such convolution can be approximated by another FSM formula
\begin{multline}
    f_{\text{conv}} (s; \mu_1, \mu_2, \sigma_1, \sigma_2, n_1, n_2) = \\
    \int \limits_{- \infty}^\infty \D s^\prime \, \fsh (s; \mu_1, \sigma_1, n_1) \fsh (s - s^\prime; \mu_2, \sigma_2, n_2) \approx \\
    \fsh (s; \mu_{\text{conv}}, \sigma_{\text{conv}}, n_{\text{conv}})
    \label{eq:fsh_conv}
\end{multline}
with parameters $\mu_{\text{conv}} (\mu_1, \mu_2) = \mu_1 + \mu_2$, $\sigma^2_{\text{conv}} (\sigma_1, \sigma_2) = \sigma_1^2 + \sigma_2^2$ and $n_{\text{conv}} (\sigma_1, \sigma_2, n_1, n_2)$, that can be determined from further physical considerations, for example matching the mean of $e^s$. 

We will find the convolution of density and speed useful momentarily, which we
now see simplifies to another $\fsh$.  Using both \eqref{eq:max_fsh_approx} and
\eqref{eq:fsh_conv}, we find
\begin{align}
    \text{C} (x; M, b, \mu, \sigma) &= \fsh (\mu, \sigma, n) * \Max^{(\log)} (M, b) \nonumber\\
    &\approx \fsh (\mu,\sigma,n) * \fsh (w; \mu_w, \sigma_w, n_w)\nonumber\\
    &\approx \fsh (x; \mu_{\text{conv}}, \sigma_{\text{conv}}, n_{\text{conv}} )
    \label{eq:fsh_max_conv}
\end{align}
where
\begin{align}
    \label{eq:mu_conv}
    \mu_{\text{conv}} (\mach, b) &= \mu + \mu_w (\mach, b) - \log 2 \\
    \label{eq:sigma_conv}
    \sigma_{\text{conv}} (b) &= \sqrt{\sigma^2 + \sigma^2_w (b)}
\end{align}
and $n_{\text{conv}} \in [ 1, \infty )$ is some yet undetermined effective convoluted shock parameter.

\subsection{Joint statistics of density and speed}\label{sec:joint_rho_v}

In what follows we revisit the statistical considerations of an ensemble emerging from snapshots of our isothermal, turbulent simulations. The primary dynamical quantities, density $\rho$ and velocity $\vec{v}$, can be treated as random variables with their own distributions. The statistical properties of each isothermal, turbulent ensemble are fully captured in the joint PDF of log density $s = \log \rho / \rho_0$ and magnitude of velocity, speed $v = | \vec{v} |$. For the purposes of this work, we use the detailed model proposed in \citet{rabatin23_correlations}, that accurately captures the statistics along with the correlations between the two quantities. The model gives an explicit formula for the joint PDF,
\begin{multline}
    f_{(s,v)}^{(V)} (s,v) \approx f_s^{(V)} (s) f_v^{(M)} (v) \, + \\
    + \, (\mathfrak{X} - 1)^{-1} \Big( e^{-s} f_s^{(K)} (s) - f_s^{(V)} (s) \Big) \Big( f_v^{(V)} (v) - f_v^{(M)} (v) \Big) ,
    \label{eq:gfun_final_V}
\end{multline}
given 6 parameters directly measured from a simulated ensemble: 1D r.m.s. Mach number $\mach$, mass-weighted Mach number $\machM = \sqrt{\langle \rho v^2 \langle / 3}$, the ratio between $\mach^2$ and its mass-weighted counterpart, $\mathfrak X = (\mach / \machM)^2$, average speed $u = \langle v \rangle$, average speed weighted by mass $u_M = \langle \rho v \rangle$, mean of log density $\mu$ and standard deviation of density, $\sigma$. These parameters, presented in Table \ref{tab:simpars}, in combination with the model presented in \citet{rabatin23_correlations} allow us to calculate any statistic involving density and speed.

\section{Thermal energy}
\label{sec:thermal}

Thermal energy, $\TE$, is derived from the equations of motion in Section \ref{sec:energy_conservation} as well as from thermodynamic considerations in Section \ref{sec:thermodynamics} General properties of the thermal energy are discussed in Section \ref{sec:thermal_general} and the PDF of thermal energy is discussed in Section \ref{sec:thermal_pdf}.

\subsection{Conservation of energy}
\label{sec:energy_conservation}

In this section, we manipulate equations (\ref{eq:hydro_cont}, \ref{eq:hydro_Euler}) in line with \citet{Banarjee18}. The goal is to obtain a conserved scalar quantity, total energy, along with its flux. We start by taking the dot product of $\vec{v}$ and Eq. \eqref{eq:hydro_Euler}, and subsequently using Eq. \eqref{eq:hydro_cont}. These steps yield the following equality
\begin{equation}
    \frac{\partial}{\partial t} \left( \frac{1}{2} \rho v^2 \right) + \divg \left( \frac{1}{2} \rho v^2 \vec{v} \, \right) = - \vec{v} \cdot \grad p
\end{equation}

Replacing $p$ with $\rho$ on the right-hand side and using the chain rule gives us
\begin{equation}
    - \vec{v} \cdot \grad p = - \divg \left( \rho \vec{v} \left( 1 + \log \rho / \rho_0 \right) \right) - \frac{\partial}{\partial t} \left( \rho \log \rho / \rho_0 \right)
\end{equation}
where $\rho_0$ is an arbitrary density scale, for example the ambient density, or, in our case, the average density, that is conventionally set to 1 in our simulations.

Putting everything together, we get the conservation of energy
\begin{equation}
    \frac{\partial \Enet}{\partial t} + \divg \vec{j}_{E, \text{net}} = 0
\end{equation}
where $\Enet = \KE + \TE$ is the total energy density per unit volume consisting of the sum of kinetic and thermal energy density per unit volume,
\begin{align}
    \KE &= \frac{1}{2} \rho v^2 \\
    \TE &= \rho \log \rho,
\end{align}
and $\vec{j}_{E, \text{net}}$ is the energy density flux per unit area and time,
\begin{equation}
    \vec{j}_{E, \text{net}} = \left[ \frac{v^2}{2} + \left( 1 + \log \rho \right) \right] \rho \vec{v}.
\end{equation}

The conservation of energy as presented here holds in the inertial subrange of the length scales of the turbulent motion, in which the energy cascades into the smaller scales and both the driving and dissipation can be neglected.

While the conservation of energy provides a useful insight into the various types of energy, in what follows we find the context for the thermal energy $\TE$ within the framework of the thermodynamical laws.

\subsection{Thermodynamical interpretation of $\TE$}
\label{sec:thermodynamics}

To get a handle on the thermal energy found in the previous section, we start from the first law of thermodynamics that holds for the LTE of an arbitrary, small volume $\D V$ within the turbulent fluid,
\begin{equation}
    \D U = T \D S - p \D V
\end{equation}
where $U$ and $S$ are internal energy and entropy contained within the volume, respectively.

The first law of thermodynamics can be re-expressed in terms of quantities defined per unit mass; $U \to u$ (internal energy per unit mass), $S \to s$ (entropy per unit mass) and $V \to 1/\rho$ (volume per unit mass). Additionally, in the isothermal framework, we can replace $p = \rho$. The first law of thermodynamics then simplifies as follows
\begin{equation}
    \D u = T \D s + \frac{1}{\rho} \D \rho
    \label{eq:internal_specific}
\end{equation}

The specific internal energy is a function of entropy and density. However, to utilize the fact, that the temperature is kept constant, a Legendre transformation introducing a thermodynamical potential involving temperature and density will prove useful. The Legendre transformation applies for the specific quantities in the standard way,
\begin{equation}
    \TEM = u - T s
\end{equation}
where $\TEM$ is the specific Helmholtz free energy (free energy per unit mass). Replacing $u$ in Eq. \eqref{eq:internal_specific} with $\TEM + T s$ results in
\begin{equation}
    \D \TEM = - s \D T + \frac{1}{\rho} \D \rho
    \label{eq:Helmholtz_specific}
\end{equation}
which shows, that $\TEM$ is a function of $T, \rho$.

This equation allows us to directly use the isothermal assumption, $T = \text{const.}$, and therefore $\D T = 0$. Integrating over $\rho$ results in
\begin{equation}
    \TEM = \log \rho / \rho_0
\end{equation}
where $\rho_0$ again serves as an arbitrary constant with units of density and is conventionally set to 1. Going back to the energy per unit volume, $\TE = \rho \TEM$, gives us
\begin{equation}
    \TE = \rho \log \rho
\end{equation}

This derivation formalizes $\TE$ within the thermodynamical framework as the Helmholtz energy density per unit volume.

\begin{figure}
\centering
\includegraphics[width=0.95\linewidth]{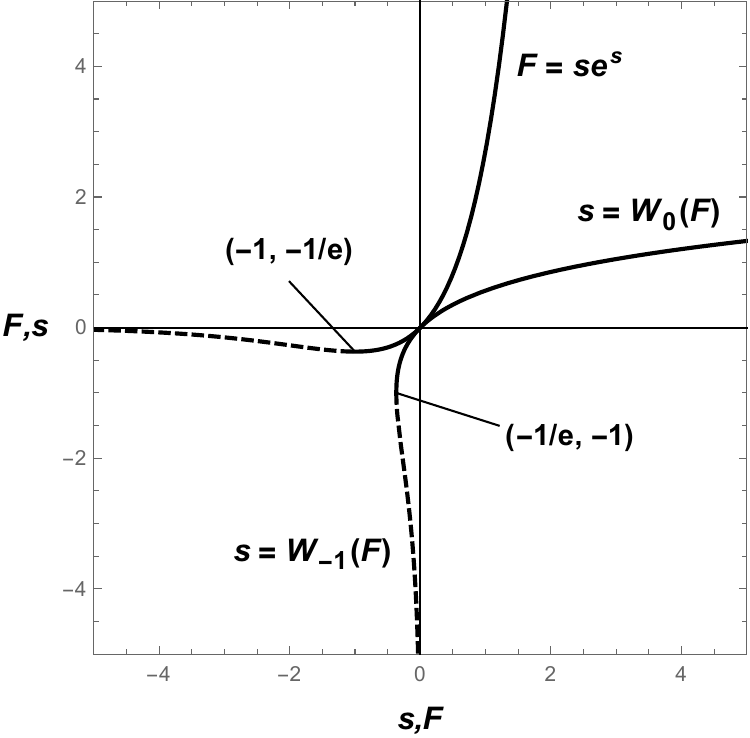}
\caption{ \label{fig:productlog} Function $\TE (s) = s e^s$ together with it's inverse, ${s (\TE) = W_{-1, 0} (\TE)}$. The branch -1 is denoted with a dashed line, while the branch 0 is plotted with solid line.}
\end{figure}

\begin{figure*} \begin{center}
    \includegraphics[width=0.99\textwidth]{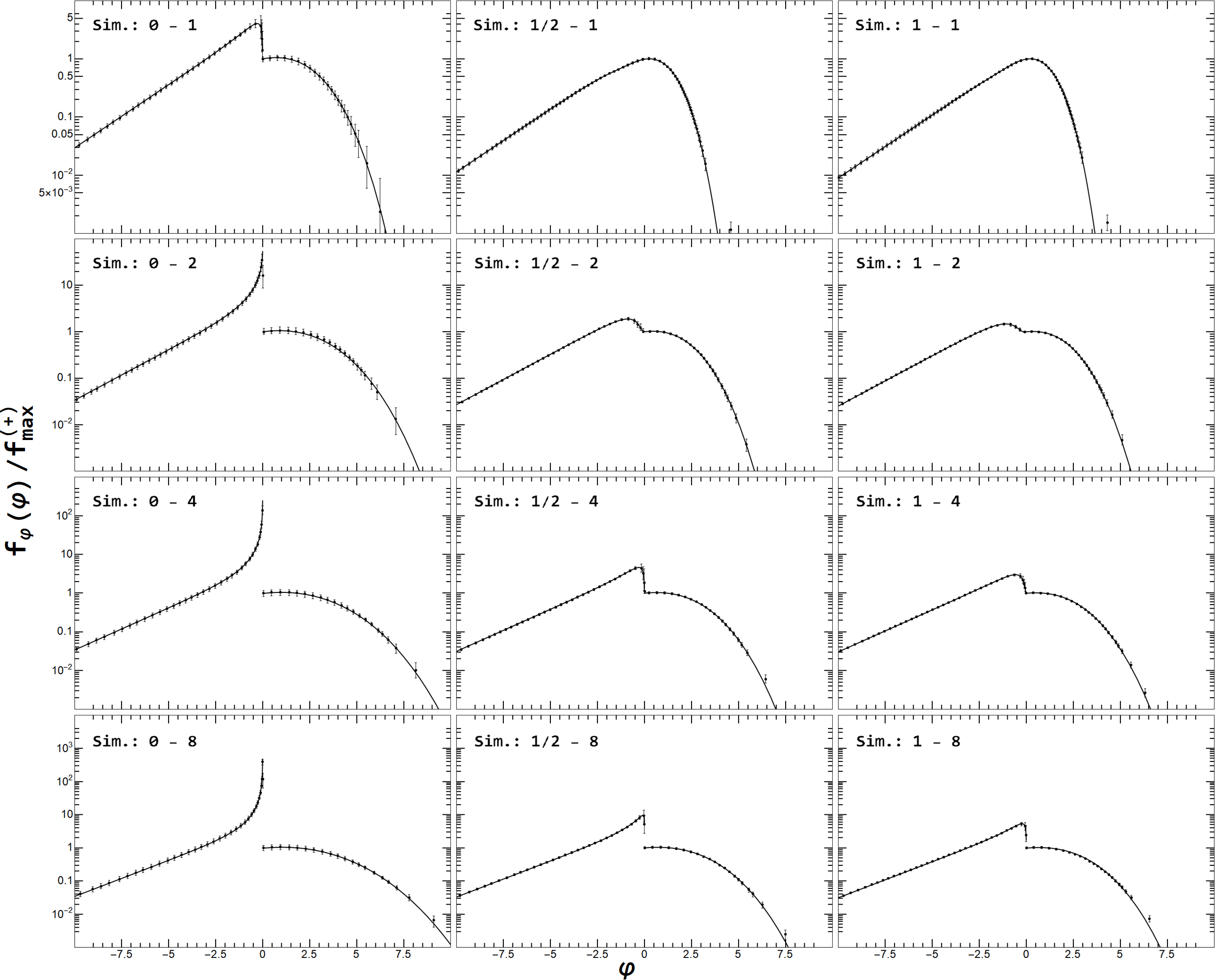}
\caption{Plots of the PDF (vertical axis) of the shifted log of thermal energy, $\te = \log (1 + e \TE)$ (horizontal axis). Dots along with horizontal bars represent data points and error bars. Solid curve depicts the exact formula \eqref{eq:log_thermal}. The PDF has been rescaled vertically by the value $\displaystyle f^{(+)}_{\text{max}} = \lim_{\varepsilon \to 0^+} f_{\varphi} (\varepsilon)$ for the sake of clarity.}
\label{fig:thermal} \end{center} \end{figure*}

\subsection{General properties, geometrical considerations}\label{sec:thermal_general}

In this section we explore the properties of the Helmholtz free energy, $\TE = \rho \log \rho = s e^s$ more closely.

The dimensionless thermal energy $\TE = s e^s$ is a non-injective function of log-density $s$ that attains its minimum $\TE_{\text{min}} = -1/e$ at $s = -1$. For $0 < s < 1$, $\TE < 0$ and is multivalued, as ${\TE (\rho=0) = \TE(\rho = 1) = 0}$.  The many-to-one property can be observed from Fig. \ref{fig:productlog}. Since the inverse function to $s e^s$ has two branches for $\TE < 0$, an additional contribution to the PDF of $\TE$ is expected for negative thermal energies. This additional weight is expected to become important should the PDF of log-density, $f_s (s)$, be broad enough to to reach both positive and negative tail in $s$. Therefore, the weight of the negative thermal energy is expected to be more prominent in high Mach numbers and/or compressive driving, as the deviation in $s$, $\sigma$, is large in both cases. Additionally, the presence of a minimum at $\TE = -1/e$ foreshadows a power law spike in the thermal energy PDF near the left edge of the domain. This feature will be used in \ref{sec:thermal_pdf} to provide an analogous random variable in a logarithmic scale, highlighting the power law behavior, $\te = \log (1 + e \TE)$. Additionally, this choice preserves the position of the singularity at $\te = \TE = 0$.

The mean value of $\TE$ can be calculated using the FSM \citep{fsh23}. Moreover, this mean value can be re-interpreted as the mass-weighted average of log density,
\begin{equation}
    \langle \TE \rangle = \langle s e^s \rangle = \langle s \rangle_M = \mu_M
\end{equation}

For ideally lognormal statistics of density, the mass-weighted mean of log-density is simply related to its volume-weighted counterpart, $\mu_M = -\mu_V = \sigma^2 / 2 > 0$. The positive sign of $\mu_M$ does not change for a finite value of $n$, therefore, the mean thermal energy is always positive $\langle \TE \rangle > 0$.


\subsection{Statistical properties}
\label{sec:thermal_pdf}

The Helmholtz free energy of the form $\rho \log \rho$, or, in terms of variable $s$, $s e^s$, is only a function of one of the primary variables - density. As such, its PDF can be obtained solely using the PDF of density by a random variable transformation and is given by equation \eqref{eq:1_1_transform} where $s, \TE$ take the role of the baseline and transformed variables, respectively.

To explicitly integrate over the $\delta$-function, the roots of equation $\TE = s e^s$ must be found for all values of $\TE \in [- 1/e, \infty)$. These roots are captured in a special function, Lambert product logarithm $W$. Since the function $s e^s$ is not monotonic, its inverse, as depicted in Figure \ref{fig:productlog}, is multivalued. For the purposes of the thermal energy PDF, we recognize two real branches of the product log, the principal branch $W_0$ for arguments $\in [-1/e, \infty)$ and values ranging from $-1$ to $\infty$, and an additional branch denoted by $W_{-1}$ for arguments $\in [ -1/e, 0)$ and values $\in (-\infty, -1]$.

With these definitions, the $\delta$-function appearing in Equation \eqref{eq:1_1_transform} can be rewritten as
\begin{equation}
    \delta (\TE - s e^s) = \sum_{k = -1, 0} \frac{\delta (s - W_k (\TE) ) \, \theta_k (\TE)}{e^s \left| 1 + s \right|}
\end{equation}
where the sum is over the two real branches of the product log and each branch only applies to its appropriate range as described earlier,
\begin{align*}
    \theta_{-1} (x) &=
    \begin{cases}
        1 ; \; -1/e \leq x < 0 \\
        0 ; \; \text{otherwise}
    \end{cases} \\
    \theta_0 (x) &=
    \begin{cases}
        1 ; \; x \geq 0 \\
        0 ; \; \text{otherwise}
    \end{cases} \\
\end{align*}

The PDF of $\TE$ is
\begin{equation}
    f_\TE (\TE) = \sum_{k = -1, 0} \frac{f_s \Big(W_k (\TE)\Big) \, \theta_k (\TE)}{\exp \Big(W_k (\TE)\Big) \left| 1 + W_k (\TE) \right|}
\end{equation}

Since the PDF of thermal energy diverges when $\TE \to -1/e$ from above, we find it advantageous to work with the PDF of its logarithmic counterpart, $\te = \log (1 + e \TE)$. The PDFs of $\te$ and $\TE$ are related by a simple variable transformation
\begin{equation}
    f_\te (\te) = e^{\te - 1} f_{\TE} (e^{\te - 1} - e^{-1})
    \label{eq:log_thermal}
\end{equation}

Fig. \ref{fig:thermal} shows the histograms of log thermal energy density obtained from the numerical simulations as solid points with error bars. Our model is depicted as a solid black curve, showing a remarkable match. The power law behavior at the left side, $\TE = -1/e$, $f_{\TE} \sim (\TE + 1/e)^\alpha$ is apparent from the linear tail as seen in plots. One curious feature is the discontinuity around $\te=\TE=0$. This feature is more pronounced for the compressively driven simulations, and for all driving modes the discontinuity increases with Mach number.  As expected, more high energy gas is created with increasing Mach number, and this effect is more pronounced with compressive driving.  Our model perfectly captures these features. 
\section{Kinetic energy}
\label{sec:kinetic}

\begin{figure*} \begin{center}
    \includegraphics[width=0.99\textwidth]{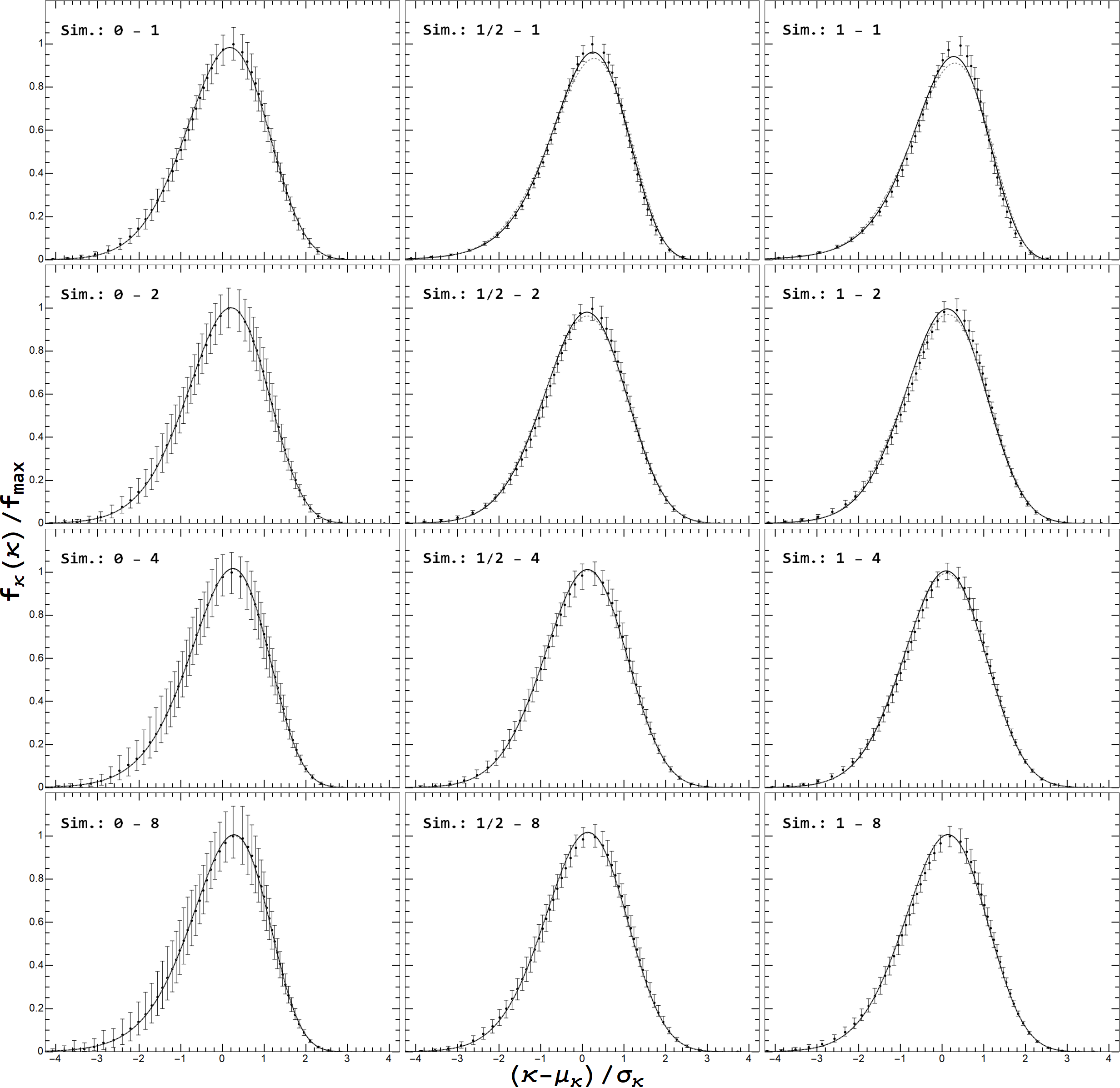}
\caption{Plots of the PDF (vertical axis) of logarithm of kinetic energy density, $\ke$ (horizontal axis). Dots and vertical lines represent data points with error bars. Solid curve depicts the exact formula using Eqs. (\ref{eq:kinetic_exact}, \ref{eq:KE_ke_transform}), dashed line represents the approximation \eqref{eq:kinetic_log_approx}. For the sake of clarity, the horizontal axis is shifted and rescaled by the corresponding mean and standard deviation, while the vertical axis is rescaled by the maximum value.}
\label{fig:kinetic} \end{center} \end{figure*}

In Section \ref{sec:energy_conservation} we showed, that the kinetic energy is $\KE = \rho v^2 / 2$. Since it depends on both density and speed, statistics of both, as well as the correlations between the two, become important.

Some general properties of the kinetic energy are discussed in Section \ref{sec:kinetic_general}. The exact formula for the PDF is discussed in Section \ref{sec:kinetic_exact}, an approximation is given in \ref{sec:kinetic_approx}.

\subsection{General properties}\label{sec:kinetic_general}

The average kinetic energy density $\langle \KE \rangle$ is directly related to one of the parameters used to construct the joint statistics of density and speed,
\begin{equation}
    \langle \KE \rangle = \frac{1}{2} \langle \rho v^2 \rangle = \frac{1}{2} \langle v^2 \rangle_M = \frac{3}{2} \machM^2
\end{equation}
where $\machM$ is the mass-weighted Mach number. For specific values, see Table \ref{tab:simpars}. It should be noted, that since density and speed are correlated, $\mach \neq \machM$ and therefore, the value for $\langle \KE \rangle$ differs from $(3/2) \mach^2$. This difference is especially prominent in the transsonic regime.

Due to the additive nature of logarithms, we will later introduce logarithmically transformed kinetic energy, $\ke = \log \KE = s + 2 \log v - \log 2$. This choice comes natural, as the result is a sum of \mbox{near-Gaussian} (albeit correlated) random variables, $s$ and $\log v$. Log-density $s$ is near-normal depending on the number of shocks, $n$, and $\log v$ is distributed according to the FSM with effective $n = 1$.

\subsection{Kinetic PDF: exact formula}\label{sec:kinetic_exact}

Following Eq. \eqref{eq:2_1_transform}, the PDF of kinetic energy density, $\KE (s, v) = e^s v^2 / 2$, can be formally written as the following transformation of random variables, given a known joint PDF of the baseline variables $s, v$
\begin{equation}
    f_{\KE} (\KE) = \int \limits_{- \infty}^\infty \D s \int \limits_0^\infty \D v \, f_{(s,v)} (s, v) \, \delta \left( \KE - \frac{1}{2} e^s v^2 \right)
\end{equation}
which can further be simplified by integrating over $v$ using the $\delta$-function
\begin{equation}
    f_\KE (\KE) = \int \limits_{- \infty}^\infty \D s \, \frac{f_{(s,v)} \Big( s, \sqrt{2 e^{-s} \KE} \Big)}{\sqrt{2 e^{s} \KE}}
    \label{eq:kinetic_exact}
\end{equation}

This exact formula cannot be simplified any further without considering a specific model for the joint statistics for density and speed, $f_{(s,v)}$. Even for the simplest case of the lognormal PDF of density and Maxwellian PDF of speed, the integral cannot be performed in a closed form.

Due to the lognormal nature of density, we examine $\ke = \log \KE$, for which the PDF is a simple variable transformation of $f_{\KE}$,
\begin{equation}
    f_\ke (\ke) = e^\ke f_\KE (e^\ke)
    \label{eq:KE_ke_transform}
\end{equation}
where $f_\KE$ is the baseline PDF of $\KE = \frac{1}{2} \rho v^2$.

\subsection{Logarithmic approximation}
\label{sec:kinetic_approx}
In this section we will derive an approximate analytic formula directly for PDF of $\ke = \log \KE$ given a few basic assumptions.

We begin by  integrating over the joint distribution,
\eqref{eq:gfun_final_V}, by way of \eqref{eq:2_1_transform}, where $R$ now
represents $\ke = \log \KE = s + w - \log 2$. Since the terms in $f_{(s,v)}$ are uncorrelated, this then becomes the convolution of five pairs of
density and speed distributions, which by \eqref{eq:fsh_max_conv} becomes
\begin{align}
    \label{eq:kinetic_log_approx}
    f_\ke (\ke) = \, & \text{C} (\ke; \machM, b_M, \mu, \sigma) \, + \\
        (\mathfrak X - 1)^{-1} \Big( \mathfrak X \, &\text{C} (\ke; \mach, b, \mu - \log \mathfrak X, \sigma) \, -\nonumber\\ 
    \mathfrak X \, &\text{C} (\ke; \machM, b_M, \mu - \log \mathfrak X, \sigma) \, -\nonumber \\
    & \text{C} (\ke; \mach, b, \mu, \sigma) \, +\nonumber \\
    &\text{C} (\ke; \machM, b_M, \mu, \sigma) \Big)\nonumber
\end{align}

Since for the general shape of the FSM formula, $\mu, \sigma, n, \langle e^s \rangle$ are tied together via eq. \eqref{eq:fsh_exp_mean}, we can re-express the mean kinetic energy $\langle \KE \rangle = \langle e^\ke \rangle = \frac{3}{2} \machM^2$, or, after multiplying by 2,
\begin{multline}
    \langle 2 e^{\ke} \rangle = 3 \machM^2 = \\
    \exp \left( \mu_{\text{conv}} (\machM, b_M) \right) \phi \left( - i \sigma_{\text{conv}} (b_M), n_{\text{conv}} \right)
    \label{eq:ke_constraint}
\end{multline}
where the cancellation of last 4 terms happens under the assumption that the dependence of $n_{\text{conv}}$ on $b$ can be neglected. Using relations (\ref{eq:muw_approx}, \ref{eq:sigmaw_approx}, \ref{eq:mu_conv}, \ref{eq:sigma_conv}), constraint \eqref{eq:ke_constraint} can be expressed as
\begin{multline}
    3 \machM^2 = \exp \left( \mu + \mu_w (\machM, b_M) \right) \times \\
    \phi \left( - i \sqrt{\sigma^2 + \sigma_2^2 (b_M)}, n_{\text{conv}} \right)
\end{multline}
This equation can be used to find $n_{\text{conv}}$ given values of $\mu, \sigma, \machM, b_M$.

Fig. \ref{fig:kinetic} shows the log kinetic energy density histograms extracted from the simulations as solid points with error bars. These are compared to our exact model (Equation \eqref{eq:kinetic_exact}, solid line) as well as the approximate 
formula (Equation \eqref{eq:kinetic_log_approx}, dashed line). It should be noted that our approximation holds even in cases where the statistics of density strongly deviates from lognormal and speed from ideal Maxwellian (see \citet{rabatin23_correlations}.) 
It deviates from the true distributions in datasets exhibiting larger correlation between density and speed, such as the transsonic simulation with mixed or purely solenoidal driving.

\section{Joint PDF}
\label{sec:joint}

\begin{figure*} \begin{center}
    \includegraphics[width=0.96\textwidth]{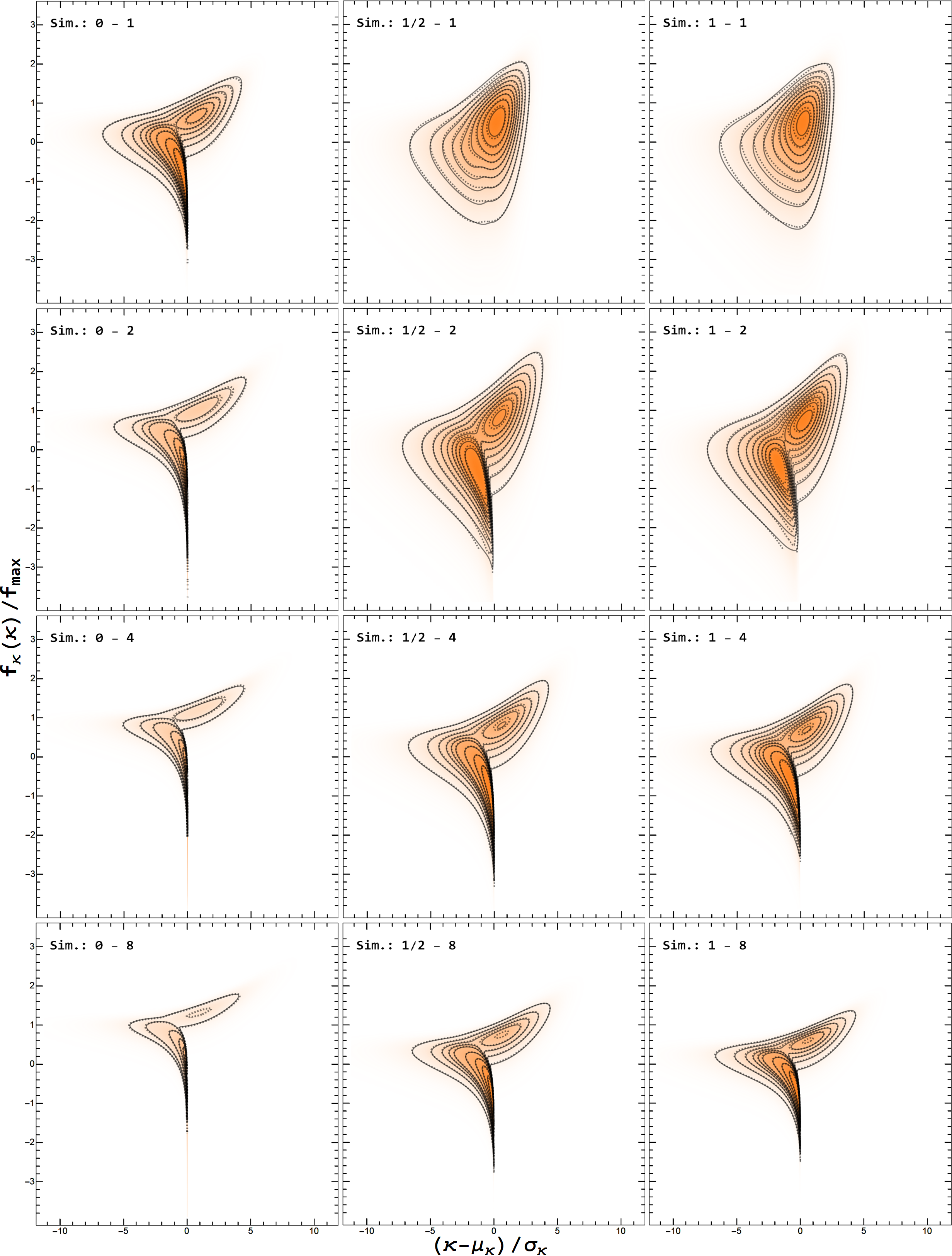}
\caption{The joint PDF of $\te$ and $\ke$ with our detailed model. Color and dashed contour shows data.  Solid line shows contours of our model. Each simulation is labeled with $\xi - M$ in the top left corner. The axes are shifted and rescaled for the sake of clarity.}
\label{fig:joint} \end{center} \end{figure*}

The joint PDF of thermal and kinetic energy can be obtained from a known joint distribution of $s, v$ and a simple random variable transform $(s, v) \to (\TE, \KE)$ using the transformation \eqref{eq:2_2_transform}.
\begin{align}
    f_{(\TE, \KE)} (\TE, \KE) &=\nonumber\\
    \int \limits_{- \infty}^\infty \D s \int \limits_0^\infty &\D v \, f_{(s,v)} (s, v) \, \delta \left( \TE - s e^s \right) \delta \left( \KE - \frac{1}{2} e^s v^2 \right)
\end{align}

Similar to the derivation of the PDF of thermal energy, we solve for the two roots $k = -1, 0$ of $s$ given a value of $\TE$ using the product logarithm, $s_k = W_k (\TE)$. The other $\delta$-function can be used to determine the sole root for $v = \sqrt{2 e^{-s} \KE}$. Together, we consider both $\delta$-functions in their root form
\begin{multline}
    \delta \left( \TE - s e^s \right) \delta \left( \KE - \frac{1}{2} e^s v^2 \right) = \\
    \sum_{k = -1, 0} \frac{\delta (s - W_k (\TE)) \delta (v - \sqrt{2 e^{-s} \KE}) \theta_k (\TE)}{e^{2 s} | 1 + s | v}
\end{multline}
and the joint energy PDF can be written analytically as follows
\begin{multline}
    f_{(\TE, \KE)} (\TE, \KE) = \\ \sum_{k = -1, 0} \frac{f_{(s, v)} \left( W_k (\TE), \sqrt{2 \exp (- W_k (\TE)) \KE} \right) \theta_k (\TE)}{\exp \left( \frac{3}{2} W_k (\TE) \right) \left| 1 + W_k (\TE) \right| \sqrt{2 \KE}}
    \label{eq:joint}
\end{multline}

Since density is distributed roughly lognormally and the thermal energy exhibits a power-law behavior near its minimum at $\TE = -1/e$ for $s = -1$, we consider the logarithm of both components of energy in the following form: $\te = \log (1 + e \TE), \ke = \log (\KE)$, for which the joint PDF can be written as
\begin{equation}
    f_{(\te, \ke)} (\te, \ke) = e^{\te + \ke - 1} f_{(\TE, \KE)} \left( e^{\te - 1} - e^{-1}, e^{\ke} \right)
    \label{eq:joint_log}
\end{equation}

The joint statistics of $\te$ (horizontal axis) and $\ke$ (vertical axis) are shown in Fig. \ref{fig:joint}. The histograms extracted from the simulations are displayed as dashed contours, while the color represents the fractional probability. Our result from equation \ref{eq:joint} transformed via \ref{eq:joint_log} is shown via solid contours. The underlying statistical model for the density-speed PDF uses the detailed basis discussed in \citet{rabatin23_correlations}. The bimodal behavior of the correlation between energies is apparent from the figure and perfectly reproduced by our model.  Each simulation has two lobes, one with high values of kinetic and thermal energies, and one with low values of kinetic and thermal energies.  Gas near the mean density rarely attains high kinetic energy.  The discontinuity at $\TE = \te = 0$ appears again, with the curious result that gas with high thermal energy must have high kinetic energy.
\section{Conclusions}
\label{sec:conclusions}

In this work we provided a deep dive into the statistics of energy in isothermal supersonic turbulence. Using a baseline model of the joint PDF of density-speed we were able to derive the marginalized PDF of thermal and kinetic energy, as well as their joint PDF.

The PDF of thermal energy carries the signature of the density-dependent thermal energy, $\TE = s e^s$. This manifests in two ways; first, a power law behavior emerges near the left edge of the thermal energy PDF, at $\TE = - 1/e$, as a direct consequence of said minimum. Second, a jump discontinuity in the PDF occurs at $\TE = 0$ due to the region $\TE < 0$ of the thermal PDF carrying additional weight originating in the non-injective nature of the thermal energy function of log-density. Specifically, the region $s \in (-\infty, -1]$ gives rise to the same values of thermal energy as $s \in [-1, 0)$. These features will show gradually with increasing width in log-density. Our model matches the numerically simulated data exceptionally well.

It is advantageous to discuss the log of the kinetic energy. The PDF of $\ke = \log \KE = s + w - \log 2$ uses the additive nature of logarithms and, provided $s$ and $w$ are independent, its PDF can be evaluated via convolution of the underlying PDFs for $s$, $w = \log v^2$. This was used with the combination with the underlying model for the joint statistics of $s, v$, consisting of 5 marginalized terms, providing an approximate formula for the PDF of $\ke$. This formula approximates the exact formula, given as a numerical integral, fairly well in all datasets except in those that have a higher degree of correlation between density and speed. The exact formula matches the data reasonably well in the same datasets, which indicates, that the correction to the joint PDF of density and speed, carrying the correlations between the two, is lacking in precision.

Finally, the joint PDF of energy was expressed in closed form and showed a remarkable match with the numerically obtained histograms. The joint PDF exhibits an interesting feature, more prominent in more supersonic simulations, that pronounces the weight of for $\TE < 0$. For highly supersonic motion, regardless of the driving pattern, most of the gas, by volume, exhibits negative thermal energy.

By exploring the statistics of energy in this simplified model of interstellar turbulence, we hope to lay groundwork for further modelling of the complex dynamics of the ISM.

\section*{Acknowledgements}
\addcontentsline{toc}{section}{Acknowledgements}

Support for this work was provided in part by the National Science Foundation
under Grant AST-1616026 and AST-2009870.  Simulations were performed on \emph{Stampede2}, part of the Extreme Science and Engineering Discovery Environment \citep[XSEDE;][]{Towns14}, which is supported by National Science Foundation grant number
ACI-1548562, under XSEDE allocation TG-AST140008.

\bibliographystyle{apj.bst}
\bibliography{main.bib}

\begin{thebibliography}{17}
\expandafter\ifx\csname natexlab\endcsname\relax\def\natexlab#1{#1}\fi

\bibitem[{{Banerjee} \& {Kritsuk}(2018)}]{Banarjee18}
{Banerjee}, S., \& {Kritsuk}, A.~G. 2018, \pre, 97, 023107

\bibitem[{{Bryan} {et~al.}(2014){Bryan}, {Norman}, {O'Shea}, {Abel}, {Wise},
  {Turk}, {Reynolds}, {Collins}, {Wang}, {Skillman}, {Smith}, {Harkness},
  {Bordner}, {Kim}, {Kuhlen}, {Xu}, {Goldbaum}, {Hummels}, {Kritsuk}, {Tasker},
  {Skory}, {Simpson}, {Hahn}, {Oishi}, {So}, {Zhao}, {Cen}, {Li}, \& {Enzo
  Collaboration}}]{Bryan14}
{Bryan}, G.~L., {Norman}, M.~L., {O'Shea}, B.~W., {et~al.} 2014, \apjs, 211, 19

\bibitem[{{Federrath}(2013)}]{Federrath13}
{Federrath}, C. 2013, \mnras, 436, 1245

\bibitem[{Federrath {et~al.}(2021)Federrath, Klessen, Iapichino, \&
  Beattie}]{Federrath21}
Federrath, C., Klessen, R.~S., Iapichino, L., \& Beattie, J.~R. 2021, Nature
  Astronomy, 5, 365

\bibitem[{{Federrath} {et~al.}(2008){Federrath}, {Klessen}, \&
  {Schmidt}}]{Federrath08}
{Federrath}, C., {Klessen}, R.~S., \& {Schmidt}, W. 2008, \apjl, 688, L79

\bibitem[{{Federrath} {et~al.}(2010){Federrath}, {Roman-Duval}, {Klessen},
  {Schmidt}, \& {Mac Low}}]{Federrath10}
{Federrath}, C., {Roman-Duval}, J., {Klessen}, R.~S., {Schmidt}, W., \& {Mac
  Low}, M.~M. 2010, \aap, 512, A81

\bibitem[{{Mac Low}(1999)}]{MacLow99}
{Mac Low}, M.-M. 1999, \apj, 524, 169

\bibitem[{{Mac Low} \& {Klessen}(2004)}]{MacLow04}
{Mac Low}, M.-M., \& {Klessen}, R.~S. 2004, Reviews of Modern Physics, 76, 125

\bibitem[{Porter {et~al.}(1999)Porter, Pouquet, Sytine, \& Woodward}]{Porter99}
Porter, D., Pouquet, A., Sytine, I., \& Woodward, P. 1999, Physica A:
  Statistical Mechanics and its Applications, 263, 263, proceedings of the 20th
  IUPAP International Conference on Statistical Physics

\bibitem[{Porter \& Woodward(2000)}]{Porter00}
Porter, D.~H., \& Woodward, P.~R. 2000, The Astrophysical Journal Supplement
  Series, 127, 159

\bibitem[{{Rabatin} \& {Collins}(2023{\natexlab{a}})}]{rabatin23_correlations}
{Rabatin}, B., \& {Collins}, D.~C. 2023{\natexlab{a}}, \mnras, 525, 297

\bibitem[{{Rabatin} \& {Collins}(2023{\natexlab{b}})}]{fsh23}
---. 2023{\natexlab{b}}, \mnras, 521, L64

\bibitem[{{Scalo} \& {Elmegreen}(2004)}]{Scalo04}
{Scalo}, J., \& {Elmegreen}, B.~G. 2004, \araa, 42, 275

\bibitem[{{Towns} {et~al.}(2014){Towns}, {Cockerill}, {Dahan}, {Foster},
  {Gaither}, {Grimshaw}, {Hazlewood}, {Lathrop}, {Lifka}, {Peterson},
  {Roskies}, {Scott}, \& {Wilkens-Diehr}}]{Towns14}
{Towns}, J., {Cockerill}, T., {Dahan}, M., {et~al.} 2014, Computing in Science
  and Engineering, 16, 62

\bibitem[{{Uhlenbeck} \& {Ornstein}(1930)}]{Ornstein}
{Uhlenbeck}, G.~E., \& {Ornstein}, L.~S. 1930, Physical Review, 36, 823

\bibitem[{{Vazquez-Semadeni}(1994)}]{Vazquez-Semadeni94}
{Vazquez-Semadeni}, E. 1994, \apj, 423, 681

\bibitem[{Woodward \& Colella(1984)}]{Woodward84}
Woodward, P., \& Colella, P. 1984, Journal of Computational Physics, 54, 115

\end{thebibliography}

\end{document}